\documentclass[aps,prd,showpacs,preprintnumbers]{revtex4}
\usepackage{graphicx}                                 
\usepackage{amsmath,amsfonts}

\usepackage{epsfig}
\usepackage{amssymb}
\usepackage{amsfonts}
\usepackage{float}
\newcommand{\Tr}{\mbox{Tr}}
\renewcommand{\Im}{\mathfrak{Im}\,}

\newcommand{\beq}{\begin{equation}}
\newcommand{\eeq}{\end{equation}}
\newcommand{\bea}{\begin{eqnarray}}
\newcommand{\eea}{\end{eqnarray}}
\newcommand{\beas}{\begin{eqnarray*}}
\newcommand{\eeas}{\end{eqnarray*}}
\newcommand{\eq}{\begin{equation}}
\newcommand{\en}{\end{equation}}
\newcommand{\eqa}{\begin{eqnarray}}
\newcommand{\ena}{\end{eqnarray}}

\begin{document}

\preprint{ITEP-LAT/2007-16, HU-EP-07/34}

\title{On the interrelation between monopoles, vortices, topological charge 
and chiral symmetry breaking: an analysis using overlap fermions for $SU(2)$}
\author{V.~G.~Bornyakov}
\affiliation{Institute for High Energy Physics, Protvino, 142281, Russia}
\affiliation{Institute of Theoretical and  Experimental Physics,
B.~Cheremushkinskaya~25, Moscow, 117259, Russia}
\author{E.-M.~Ilgenfritz} 
\affiliation{Humboldt-Universit\"at zu Berlin, Institut f\"ur Physik, 
Newtonstrasse~15, 12489 Berlin, Germany}
\author{B.~V.~Martemyanov}
\affiliation{Institute of Theoretical and  Experimental Physics, 
B.~Cheremushkinskaya~25, Moscow, 117259, Russia}
\author{S.~M.~Morozov}
\affiliation{Institute of Theoretical and  Experimental Physics, 
B.~Cheremushkinskaya~25, Moscow, 117259, Russia}
\author{M.~M\"uller-Preussker}
\affiliation{Humboldt-Universit\"at zu Berlin, Institut f\"ur Physik, 
Newtonstrasse~15, 12489 Berlin, Germany}
\author{ A.~I.~Veselov}
\affiliation{Institute of Theoretical and  Experimental Physics, 
B.~Cheremushkinskaya~25, Moscow, 117259, Russia}

\date{February 27, 2008}

\begin{abstract}
We study the properties of configurations from which P-vortices 
on one hand or Abelian monopoles on the other hand have been removed.
We find that the zero modes and the band of non-zero modes close to
zero disappear from the spectrum of the overlap Dirac operator,
confirming the absence of topological charge and quark condensate.
The different behavior of the modified ensembles under smearing 
compared to the unmodified Monte Carlo ensemble corroborates these 
findings. The {\it gluonic} topological susceptibility rapidly 
approaches zero in accordance with $Q_{index}=0$.
The remaining (ultraviolet) monopoles without vortices and -- to a 
less extent -- the remaining vortices without monopoles are unstable 
under smearing whereas smearing of the unmodified Monte Carlo ensemble
effects the monopoles and vortices only by smoothing, reducing
the density only slightly.  
\end{abstract}


\maketitle

\section{Introduction}
\label{sec:Introduction}

There are two popular phenomenological scenarios explaining confinement in 
lattice gluodynamics, monopole condensation~\cite{monopole} and the center 
vortex~\cite{greensite_1} mechanism. The basic ideas of both scenarios go
back to 't Hooft~\cite{tHooft_1,tHooft_2}.
Both have been recently discussed in Refs.~\cite{Greensite-review,Greensite-Alkofer}.
For our discussion, monopoles and center vortices are defined by projection 
to Abelian $U(1)^{N-1}$ or $Z(N)$ gauge fields, respectively. 
In order to distinguish them from extended vortices, these $Z(N)$ center 
vortices are called P-vortices. 
The two types of excitations, when derived from the Maximally Abelian Gauge (MAG) 
or from the Maximal Center Gauge (MCG), respectively, reproduce about 
90\%~\cite{simann} and about 70\%~\cite{drama} of the non-Abelian string tension.
This observation is called monopole and center dominance.
The monopole dominance should not be confused with Abelian dominance, which 
describes the fact that the projected degrees of freedom (the $U(1)^{N-1}$ valued links)
reproduce the original string tension equally well. Without gauge fixing the full
static potential is reproduced by abelian projected or center projected 
links~\cite{Ogilvie,Faber-Greensite-Olejnik}.

The importance of the topological excitations and of the corresponding MAG or 
MCG fixing, rests more on the physical reality of the excitations as the 
possibly relevant infrared degrees of freedom than on the monopole or P-vortex 
dominance. The reality is witnessed by their localization and the local excess 
of action and topological charge carried by monopoles and P-vortices. 
The infrared degrees of freedom could be used to derive effective theories 
to describe the infrared physics, for instance the Dual Ginzburg-Landau 
theory~\cite{DGL}. This letter elaborates on some other aspects of the physical 
reality of monopoles and P-vortices. It turns out that they are constitutive 
also for other nonperturbative features besides confinement.
Removing monopole degrees of freedom~\cite{miyamura_1,miyamura_2} or 
P-vortices~\cite{deforcrand,solbrig,morozov} from the (lattice) fields should 
leave only inert and topologically trivial gauge field configurations.

For the issue of physical reality the conjecture~\cite{conjecture,greensite_4} 
was very important that monopoles and P-vortices are geometrically interrelated. 
Indeed, this was found to be the case in $SU(2)$ gluodynamics. More than 90\% of 
monopole currents are localized on the 
P-vortices~\cite{greensite_4,interplay,kovpolsyrzakh}. The effect of eliminating
one or the other, however, is more complicated and obviously destroys this 
geometrical interrelation. 

It was realized that the removal 
of monopoles destroys only large (infrared) P-vortex clusters whereas the total 
density of P-vortex plaquettes is suppressed by less than an order of 
magnitude~\cite{mmp}. In the case of removal of vortices the total density of 
monopole links is even increased compared to the initial equilibrium 
configurations~\cite{mmp}. In that paper we have confirmed (for a finite temperature 
$T \approx 0.75~T_{\rm dec}$ in the confinement phase) that for the manipulated 
lattice ensembles confinement is missing. In particular, we were able to point out
why the apparently percolating clusters of monopoles remaining after vortex 
removal cannot produce confinement. After this observation it is impossible to 
directly infer confinement from the existence of percolating monopoles. 

In the present paper we turn our attention to the topological and chiral aspects
of monopoles and vortices. 
The first new element in comparison with all previous studies of monopole removal 
and all but one paper on vortex removal is that
we use chirally perfect overlap 
fermions~\cite{Neuberger1,Neuberger2} as a probe to confirm the loss of topological 
charge and the vanishing of the quark condensate in the modified ensembles 
obtained by monopole or vortex removal. 
The other new feature is that for the first time we employ an improved lattice 
gauge action in this kind of studies.
For all 50 configurations in the original 
and the modified ensembles we have determined 20 (in modulus) lowest eigenvalues
$\lambda_N$ and the corresponding eigenmodes (not used here) of the massless 
Neuberger overlap Dirac
operator. To identify the topological charge we refer to the index of this Dirac 
operator. This is complemented by an improved gluonic expression for the topological
density, leading to the same conclusion. We notice that the gluonic measurement 
of the topological charge requires a certain amount of APE smearing. This gives 
us the opportunity to investigate in what respect the modified ensembles differ 
from the original Monte Carlo one with respect to moderate smearing. For this part 
of our study an enlarged ensemble of 100 configurations has been used. 

In section~\ref{sec:setup} we give necessary information about the lattice
ensembles that we have used. In section \ref{sec:spectrum} the effect of the 
removal of monopole, photon and vortex degrees of freedom on the spectrum 
close to $\lambda_N=0$ and on the topological charge is described.
In section \ref{sec:smearing} we describe the behavior under
smearing - concerning the measured monopole, vortex and topological content -
which is strikingly different between the equilibrium ensemble and the modified 
ensembles with monopoles and vortices removed. 
Section \ref{sec:conclusions} contains our conclusions. In the Appendix
all necessary definitions are collected.

\section{Simulational setup}
\label{sec:setup}

In two previous papers~\cite{mmp1,luschevskaya} we have applied the overlap 
Dirac operator for $SU(2)$
lattice gauge theory in conjunction with the tree-level tadpole-improved
Symanzik action. This will be the setup also here. The overlap 
construction~\cite{Neuberger1,Neuberger2} provides a perfectly chiral description 
for lattice fermions. The choice of action is motivated as follows.
In our first paper~\cite{mmp1} we have applied the overlap Dirac operator for a 
very specific investigation, to find evidence for a partially dyonic, partially 
caloronic structure of the topological charge distribution at $T=T_{\rm dec}$. 
For this purpose it was essential to make sure that the configurations are smooth 
enough such that the number of zero modes and the gross structure of the spectrum 
of lowest overlap Dirac eigenvalues are robust with respect to a change of temporal 
boundary conditions and with respect to smearing. This would not be the case for 
the Wilson action. For the tree-level tadpole-improved Symanzik at high enough 
$\beta_{\rm imp}$ this requirement is fulfilled. This has determined us to work 
on a $20^3\times 6$ lattice in the paper~\cite{mmp1} where 
$\beta_{{\rm imp},c}=3.25$ has been found to be the deconfinement critical point.
In a second paper~\cite{luschevskaya} we have extended our investigation with
this action to temperatures $T$ below $T_c$ and and up to $2~T_c$. Here our
focus was the dependence of the spectral density and the localization behavior
of the eigenmodes on the sign of the spatially averaged Polyakov loop $L$
as soon as it ceases to vanish in the deconfined phase. In this paper it has been 
found that a gap opens in the spectral density for $T  > 1.05~T_{\rm dec}$, but 
only for configurations with a positive spatially averaged Polyakov loop, $L > 0$, 
in agreement with predictions by Stephanov~\cite{stephanov}.

We refer to these two papers for details concerning the action and the 
implementation of the overlap Dirac operator.
For the present investigation we have chosen the same lattice size $20^3\times 6$ 
and the same action. We work at $\beta_{{\rm imp},c}=3.25$. There is no need to 
compare different boundary conditions. Now the overlap Dirac operator is uniquely 
endowed with antiperiodic boundary conditions in the temporal and periodic ones 
in the spatial directions. We have extended the ensemble to 50 configurations, 
that was begun with 20 configurations for Ref.~\cite{mmp1}. 
The eigenvalues $\lambda_{\rm imp}$ of the improved Neuberger 
operator~\cite{improved_overlap} are obtained by stereographic projection 
from $\lambda_N$ situated on the Ginsparg-Wilson circle 
onto the imaginary axis, such that in the following the eigenvalues are 
understood as $\lambda \equiv \Im \lambda_{\rm imp}$. 

Since we have chosen to work at the deconfinement temperature, it makes no sense
to discuss here once more the influence of the removal of monopoles or vortices 
on the string tension~\cite{mmp}. We are concentrating here on the effect of
monopoles and vortices on the topological charge $Q$ via the Atiyah-Singer 
index theorem~\cite{AtiyahSinger},
\begin{equation}
Q = N_{-} - N_{+} \; ,
\end{equation}
with $N_{+}$ and $N_{-}$ the number of zero modes of positive and negative 
chirality, and the spectral density $\rho(\lambda)$ near $\lambda=0$, which is 
related via the Banks-Casher relation~\cite{BanksCasher} 
\begin{equation}
\langle \bar{\psi}\psi \rangle = - \frac{\pi~\rho(0)}{V} 
\end{equation}
(with the four-volume $V = N_s^3~N_t~a^4$) 
to the quark condensate. Whenever more than one zero mode is found for one
configuration, the chirality of all of them is found to be the same.

For the ensemble of configurations at $T=T_{\rm dec}$ in Ref.~\cite{mmp1} the 
gross spectral density was seen to be independent of the boundary condition 
which means that it is insensitive also to the sign of the (very small) 
average Polyakov loop in our ensemble. This makes it possible to study the 
non-trivial effect of monopole or vortex removal on the spectral density in 
an unambiguous way.

The monopole line density~\cite{monopoledensity} and the P-vortex plaquette 
density referred to later are defined in units of the lattice spacing as
\begin{equation}
\rho_{\rm mon}~a^3 = \frac{<N_{\rm mon}>}{4 N_s^{3} N_{t}} \quad \mathrm{and} \quad
\rho_{\rm vort}~a^2 = \frac{<N_{\rm vort}>}{6 N_s^{3} N_{t}} \; ,
\label{eq:density_def}
\end{equation}
where $N_{\rm mon}$ is the number of dual links carrying non-vanishing monopole 
currents for an Abelian projected gauge field obtained from the MAG. 
$N_{\rm vort}$ is the number of dual plaquettes belonging to the total P-vortex 
area after applying the center projection to the gauge field put into the direct
maximal center gauge (DMCG)~\cite{DMCG}. More about these definitions and the 
procedures that lead to the detection and removal of monopoles and vortices 
can be found in the Appendix.  

The following results concerning the density of monopoles and vortices
and the investigation of the behavior under smearing in section
\ref{sec:smearing} are based on an ensemble of 100 configurations.

At $T=T_{\rm dec}$, for our equilibrium configurations the vortex density is 
found to be $\rho_{\rm vort}~a^2 = 0.0231(4)$. With the zero-temperature 
string tension $\sigma_0 = \sigma(T = 0)$ setting the scale, this corresponds 
to $\rho_{\rm vort}/\sigma_0 = 0.417(8)$ or 
$\rho_{\rm vort} = 2.08(4) {\rm~fm}^{-2}$.
This number is essentially smaller than the zero temperature density found
with the Wilson action to be about $4 {\rm~fm}^{-2}$ \cite{gubarev}. 
We think that this large difference is mostly due to difference in actions 
used in Ref.~\cite{gubarev} and in this paper rather than due to finite 
temperature effects, i.e. it indicates that the vortex density with improved 
action is substanially smaller than with Wilson action.

The monopole density in the equilibrium ensemble amounts to 
$\rho_{\rm mon}~a^3 = 0.0117(1)$, that means
$\rho_{\rm mon}/\sigma_0^{3/2} = 0.897(9)$ or 
$\rho_{\rm mon} = 10.0(1) {\rm~fm}^{-3}$.
For comparison, we recall an estimate~\cite{universality} of the monopole 
density at $T=0$ for $\beta_{\rm imp} = 3.25$ and the same action:
$\rho_{\rm mon}~a^3 = 0.0126(1)$. That means that the monopole density 
is only insignificantly suppressed at $T_{\rm dec}$ compared to $T=0$.

These densities have to be taken with a grain of salt because it is known
that they should be decomposed in an infrared and an ultraviolet part.
The presence of both components becomes obvious in studies of 
universality~\cite{universality} where only the infrared part possesses
a finite continuum limit that can be compared between different actions.
This problem will show up in the process of smearing discussed in 
section~\ref{sec:smearing}.

The topological susceptibility of the equilibrium ensemble defined by 
$\langle Q_{\rm overlap}^2 \rangle = 7.3 \pm 1.5$
translates to $\chi_{\rm top}/\sigma_0^2 = 0.049 \pm 0.010$  
or, assuming $\sqrt{\sigma_0}=440 {\rm~MeV}$, 
to $\chi_{\rm top} = (207 \pm 10 {\rm~MeV})^4$. This is in the right ballpark 
for the (unsuppressed) topological susceptibility. Once monopole or vortex
degrees of freedom are removed from the configurations the fermionic (overlap) 
topological charge is stricly vanishing. It might come unexpectedly, that
$Q_{\rm overlap}$ is not preserved if the regular (photon) degrees of freedom
are removed from the configurations. We will point out later that in this case
also the opposite (vanishing topological charge) could have been expected. 
For this modified ensemble $\langle Q_{\rm overlap}^2 \rangle = 5.3 \pm 1.25$  
has been measured.

\section{The spectrum of low-lying modes for configurations with removed 
monopoles, photons and vortices}
\label{sec:spectrum}

In this section we demonstrate the disappearance of the 
quark condensate and the complete loss of topological charge in modified 
ensembles of configurations having monopoles or vortices removed. 
We emphasize that this effect has been partly studied 
already~\cite{miyamura_1,miyamura_2,deforcrand,solbrig,morozov}, 
for either monopoles or vortices removed, with various gauge actions 
and fermionic actions (staggered, chirally improved) for the confinement phase. 
Thus, the results of this section are a confirmation of the crucial role of 
monopole and vortex degrees of freedom for the spectral properties 
obtained with a new, more convincing tool - the overlap Dirac operator.
Let us note that results obtained at the edge of the confining phase of 
quenched $SU(2)$ gluodynamics are not less interesting. At this temperature 
the topological susceptibility is still approximately the same as for $T=0$, 
and the spectral gap is not yet opened~\cite{luschevskaya}.

\begin{figure}[!t]
\vspace{1 cm}
\begin{center}
\includegraphics[width=0.70\textwidth,height=.80\textwidth,angle=270]{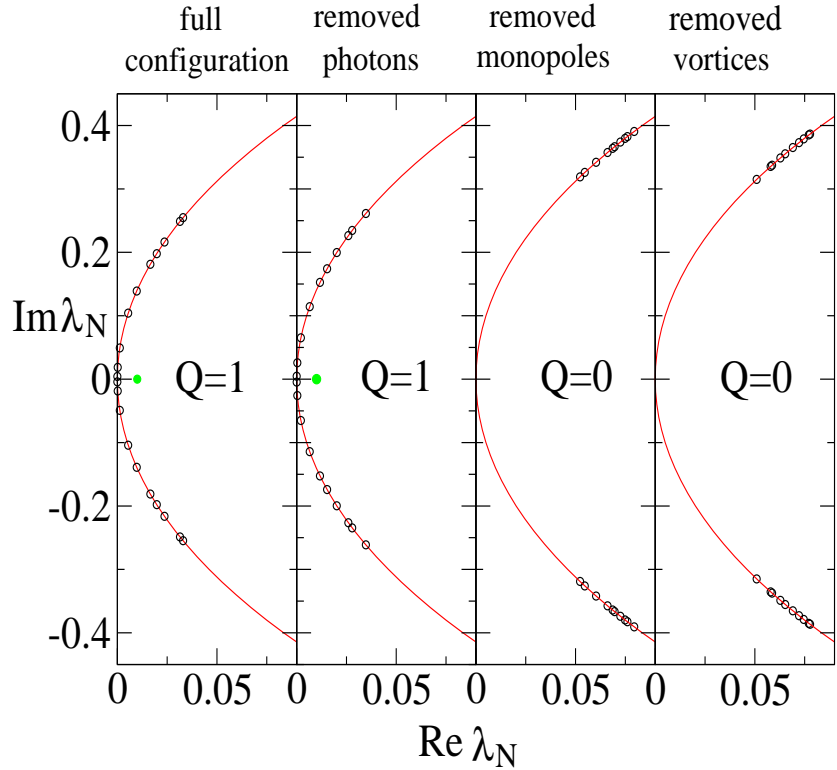}\\
\caption{The eigenvalues of the 20 lowest non-zero modes (open symbols) and the 
eventual zero mode (filled symbol) of the overlap Dirac operator lying on the 
Ginsparg-Wilson circle for one of the equilibrium configurations; for 
the original configuration (extreme left), 
the configuration with removed photon degrees of freedom (middle left),
with removed monopole degrees of freedom (middle right) and with 
vortices removed (extreme right). The zero mode (green in online color) is pulled 
away from the Ginsparg-Wilson circle for better visibility of all modes.}
\label{fig:fullphotmonvor}
\end{center}
\end{figure}

In Fig.~\ref{fig:fullphotmonvor} we illustrate this by one configuration.
The panel on the extreme right shows the spectrum of low lying modes of overlap 
fermions after removing the vortex degrees of freedom. This should be compared 
with the original spectrum shown on the extreme left. The originally existent 
zero mode has disappeared, and the non-zero modes have moved outward.
A similar change of the spectrum can be observed on the middle right panel after 
removing the monopole degrees of freedom from the same configuration. 
In the middle left panel the effect of removing the regular (photon) part from 
the Abelian projected field is shown. This spectrum differs only in minor details 
from the original spectrum but the number of zero modes and the interval covered 
by the 20 modes remained unaffected. We have to stress that the number of zero
modes is not always stable with respect to the removal of the regular part of
the Abelian field. The latter comparison enforces the conclusion that, from the 
point of view of Abelian projection, the decisive role for chiral symmetry 
breaking is played by the monopole part of the Abelian projected field, whereas
the topological charge is not robust against the reduction of the Abelian field
to its singular (monopole) part. 

\begin{figure}[!htb]
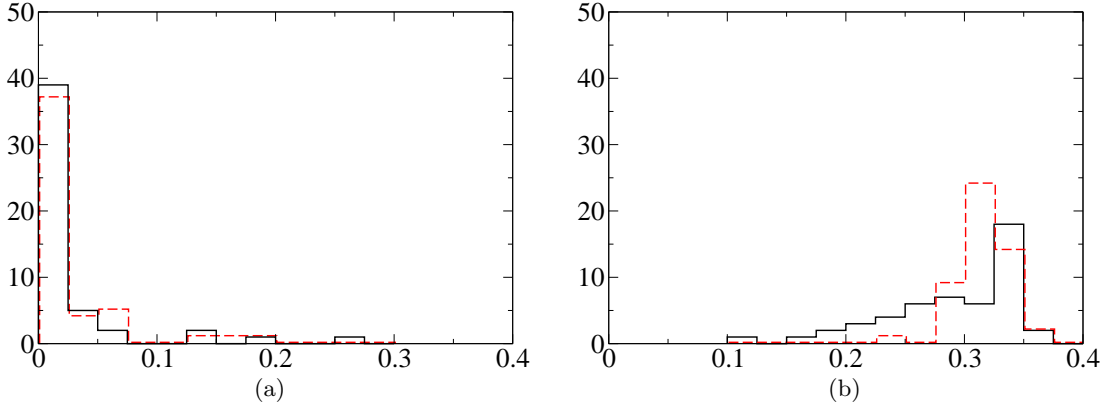

\begin{center}
\includegraphics[width=.45\textwidth,angle=0]{gap1.eps}%
\hspace{.5 cm}
\includegraphics[width=.45\textwidth,angle=0]{gap2.eps}\\
\hspace{0.0 cm} (a) \hspace{7.0 cm} (b) \\
\caption{The gap distributions (distributions of the first non-zero positive
eigenvalue $\lambda_1$) for the four ensembles of 50 lattice configurations:
(a) the original equilibrium configurations (solid line)
and the configurations with removed photons (dashed line)
(b) the configurations with removed monopoles (solid line) 
and with removed vortices (dashed line).}
\label{fig:gap}
\end{center}
\end{figure}
In Fig.~\ref{fig:gap} we show the distribution of 
the first non-zero, positive eigenvalue $\lambda$ for the four ensembles. 
After removing the regular part of the
Abelian projected gauge field the distribution is only very minimally
widened compared to the distribution of the original ensemble 
(see Fig.~\ref{fig:gap} (a) ).
After removing monopoles the distribution changes completely. It is rather 
wide with some tail towards $\lambda=0$, but with a clear gap separating it
from $\lambda=0$, while the gap is wider and less fluctuating in the case of 
removed vortices (see Fig.~\ref{fig:gap} (b) ).

\begin{figure}[!t]
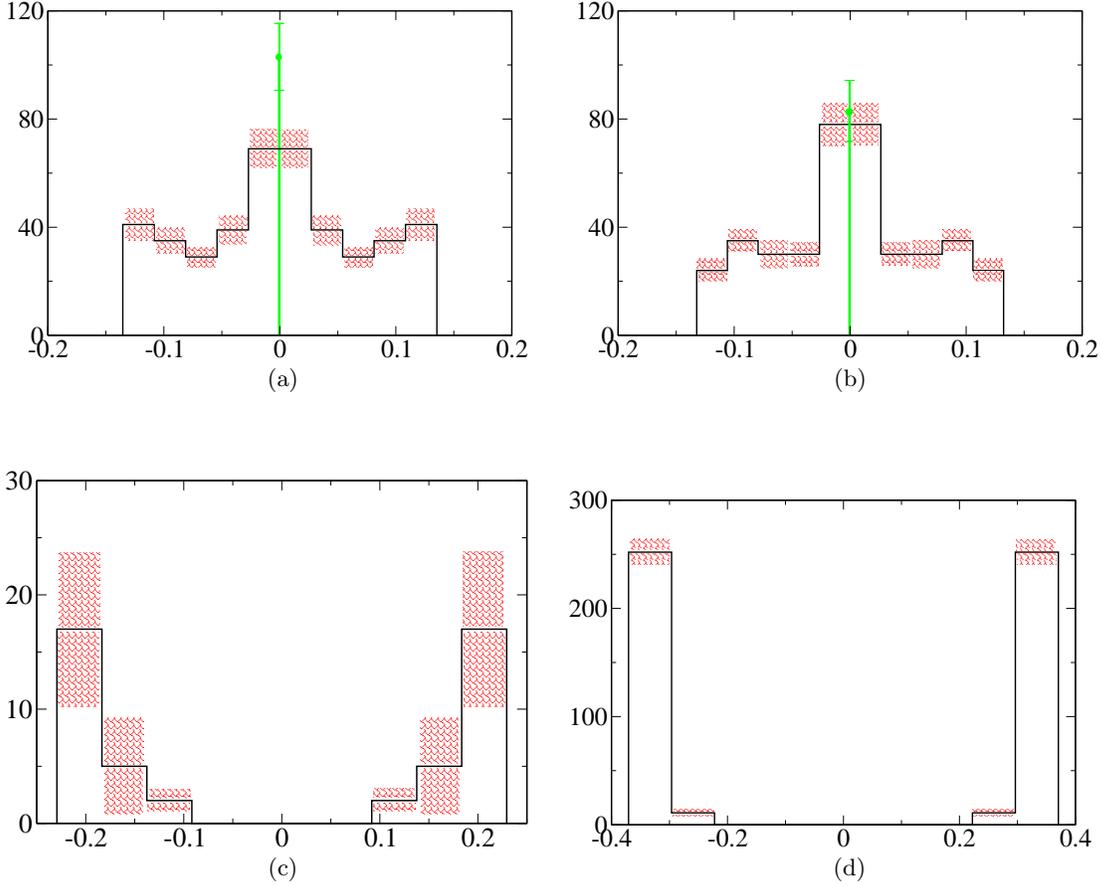

\begin{center}
\includegraphics[width=.45\textwidth,angle=0]{spektr1.eps}%
\hspace{.5 cm}
\includegraphics[width=.45\textwidth,angle=0]{spektrphot1.eps}\\
\hspace{0.3 cm} (a) \hspace{6.9 cm} (b) \\
\vspace{1.0 cm}
\hspace{-0.2 cm}
\includegraphics[width=.45\textwidth,angle=0]{spektrmon1.eps}%
\hspace{.4 cm}
\hspace{-0.1 cm}
\includegraphics[width=.45\textwidth,angle=0]{spektrvor1.eps}\\
\hspace{0.3 cm} (c) \hspace{6.9 cm} (d) \\
\caption{The histogram of the eigenvalues $\lambda$ for 20 lowest non-zero 
modes. The shadowing on top of the columns (red in color online) represents 
the statistical error. 
The total number of zero modes of the overlap Dirac operator is shown by the 
height of the spike (green in color online) drawn at zero.
The plots show the four corresponding ensembles of 50 lattice configurations:
(a) the original equilibrium configurations, 
(b) the configurations with removed photons,
(c) the configurations with removed monopoles and 
(d) with removed vortices.
The spectra are shown in an interval defined by the minimum over 50 configurations 
of the largest (in modulus) of 20 individual configuration eigenvalues, thus 
eliminating the dependence of the shown part of the spectrum on the number 
of actually calculated eigenvalues.}
\label{fig:histo}
\end{center}
\end{figure}

The respective average cumulated spectral densities as anticipated from all 
lowest 20 eigenmodes in our equilibrium
ensemble of 50 configurations and the modified ensembles are shown as 
histograms in Fig.~\ref{fig:histo}.
The spike at zero shows the total number of zero modes in the ensemble of 50 
configurations (irrespective of their chirality). The height corresponds to 
the scale of the embedding histogram. We see that zero modes and hence the
topological charge completely disappear after monopoles (bottom left) or 
vortices (bottom right) are removed from the configurations while the spectral
density is pushed outward. This effect is stronger if vortices are removed, 
less if monopoles are removed. The appearance of the gap among the near-zero 
modes signals the vanishing of the quark condensate which is still 
non-vanishing at the given temperature for the original ensemble (top left) 
and after the removal of the photon degrees of freedom (top right). 

Notice that in all configurations the zero modes are neither completely preserved 
nor completely destroyed when the photon degrees of freedom are suppressed. 
All we can say here is that the number of zero modes is a less robust feature 
provided only the monopole degrees of freedom are kept.
The numbers of zero modes in the equilibrium configurations, $N_{\rm equil}$, 
and in the
corresponding no-photon configurations, $N_{\rm no~phot}$, are strongly correlated.
From the scatter plot of both numbers a regression formula
$N_{\rm equil} = 0.15(25) + 0.77(10) N_{\rm no~phot}$ can be extracted.
This finding might be difficult to reconcile with the notion of Abelian 
dominance of the topological charge~\cite{dyons,miyamura_3}. 
Strict Abelian dominance of the latter
would imply that after the removal of the regular ``photon'' part of the
Abelian projected gauge field no topological charge should be left at all. 
These considerations were refering, however, to cooled configurations and
the ground state whereas we consider here unsmeared configurations at
non-zero temperature.   

\section{Smearing of configurations void of monopoles or vortices}
\label{sec:smearing}

We can describe the interrelation between monopoles and P-vortices 
by the effect of removing one type of infrared degrees of freedom on 
the density of the other. We should emphasize again that the density 
alone does not decide about the confining property of the 
ensemble~\cite{mmp}.

Here we are asking whether smearing is able to reveal that a lattice 
ensemble is corrupted in an essential way by the removal of monopoles
or vortices. In the confinement phase ``corrupted'' means that it is 
unable to confine. We answer this question affirmatively without 
reference to the string tension showing that the density of complementary 
objects and the topological susceptibility disappear under smearing. 
We have seen already that the effect of vortex or monopole removal on 
the chirally perfect Dirac spectrum shows up without smearing. This 
does not change afterwards under the influence of smearing.

\subsection{Vortices removed}

Let us now consider the effect of removing P-vortices from the 
configurations performing the link operation (\ref{eq:vremoved})
on the monopole content and on the topological charge. 
At this step, the monopole density $\rho_{\rm mon}$ is 
approximately doubled to $\rho_{\rm mon}~a^3 = 0.0233(3)$. 
The abundant monopole lines form clusters that still contain a 
percolating component, but the non-Abelian string tension vanishes 
as it should~\cite{deforcrand}. Also the monopole string tension 
vanishes~\cite{mmp} in agreement with expectations. The unphysical 
(inert) character of magnetic monopoles in the configurations modified 
by vortex removal was thoroughly discussed in~\cite{mmp}. In essence, 
the monopole clusters were found to be decomposable into small monopole 
loops, such that the magnetic currents are screened at large distances. 

\begin{figure*}[!t]
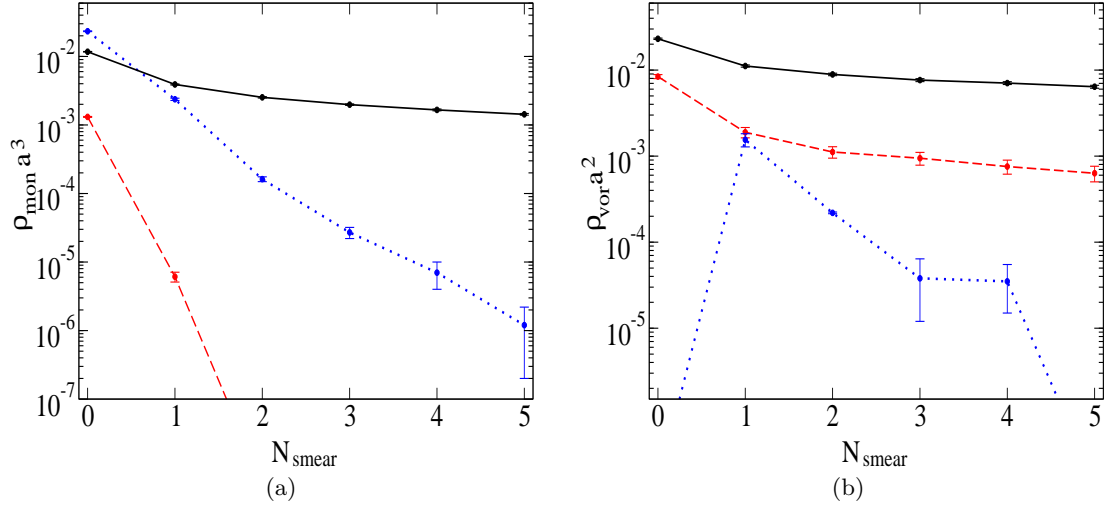

\vspace{1 cm}
\begin{center}
\includegraphics[width=.45\textwidth,height=.4\textwidth]{new_mon_lw.eps}%
\hspace{.5 cm}
\includegraphics[width=.45\textwidth,height=.4\textwidth]{new_vor_lw.eps}\\
\hspace{0.2 cm} (a) \hspace{6.9 cm} (b)
\vspace{-0.2cm}
\caption{The monopole density (a) and the vortex density (b)
depending on the number of smearing steps for the original equilibrium
ensemble of 100 configurations (solid lines) and for the modified ensembles 
differing by the removal of P-vortices (dotted lines) and the removal of 
monopoles (dashed lines).}
\label{fig:smearmonvortdens}
\end{center}
\vspace{-0.2cm}
\end{figure*}

\begin{figure}[!htb]
\vspace{1 cm}
\begin{center}
\includegraphics[width=.55\textwidth,height=.4\textwidth]{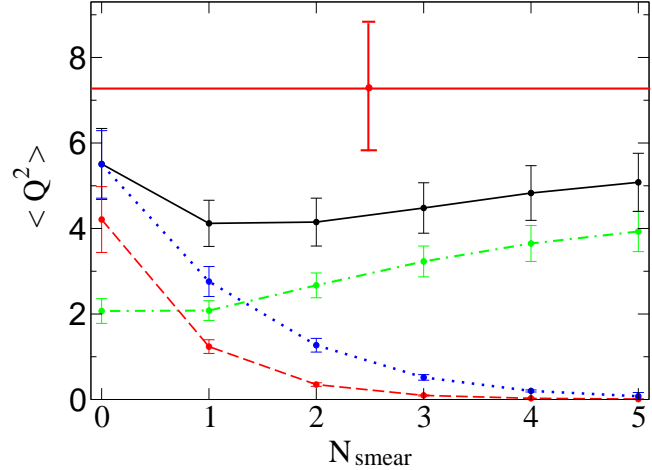}\\
\vspace{-0.2cm}
\caption{The average of the gluonic topological charge squared, 
$\langle Q_{\rm gluonic}^2 \rangle$, depending on the number of smearing steps 
for the original equilibrium ensemble of 100 configurations (solid line) 
and for the modified ensembles differing by the removal of P-vortices 
(dotted line), the removal of monopoles (dashed line) and the removal of 
photon degrees of freedom (dash-dotted line). The horizontal solid line above 
(with the statistical error bar) shows  $\langle Q_{\rm overlap}^2 \rangle$ 
for the unmodified equilibrium ensemble of 50 configurations (red in color online).} 
\label{fig:smeartopcharge}
\end{center}
\vspace{-0.2cm}
\end{figure}

The effect of vortex removal is presented by dotted lines  
in Fig.~\ref{fig:smearmonvortdens} (a) for the monopole density,
in Fig.~\ref{fig:smearmonvortdens} (b) for the vortex density
and in Fig.~\ref{fig:smeartopcharge} for the topological susceptibility.

The monopole density is initially even enhanced compared to the unmodified 
ensemble before it is quickly wiped out by smearing. In contrast to this, 
the monopole density in the unmodified ensemble is only slowly reduced 
by smearing (only one order of magnitude within 5 steps). This effect of
smearing reflects mainly the elimination of ultraviolet monopole objects. 
These are small monopole loops that are either appended to large loops 
or separately existing. The extended infrared monopole clusters 
survive with ultraviolet loops stripped off. This can be called ``smoothing of 
monopoles''.  

The effect of vortex removal on the vortex density is demonstrated 
by a dotted line  in Fig. \ref{fig:smearmonvortdens} (b).
The vortex density reappears after one smearing step at a very low level 
before it is finally rapidly wiped out by smearing. In contrast 
to this, the vortex density in the unmodified ensemble is slowly reduced 
by smearing (only by a factor of three within 5 steps). This effect of
smearing reflects mainly the straightening of the vortex surface
due to elimination of ultraviolet objects (these are isolated bubbles and 
decorations added to extended surfaces) whereas large infrared objects survive. 

In Fig. \ref{fig:smeartopcharge} the effect of vortex removal on the 
topological susceptibility as determined by the gluonic topological
density is also shown by a dotted line as function of the smearing
steps.  More precisely, the susceptibility is quantified by the average 
of the topological charge squared $\langle Q_{\rm gluonic}^2 \rangle$. 
In contrast to the original ensemble, where the topological
susceptibility slowly approaches some final value from below, in
the ensemble without vortices the (gluonic) topological susceptibility
decays to zero within only 5 smearing steps. 

For the gluonic topological charge density the topological susceptibility 
is known to receive additive and multiplicative 
renormalization~\cite{DiGiacomo}. Already few smearing steps show that the 
gluonic topological susceptibility becomes rapidly readjusted to zero in 
the modified ensemble without vortices for which the index of the overlap 
Dirac operator gives $Q_{\rm overlap}=0$ right from the beginning (before 
smearing starts). For this case the dotted curve in Fig.~\ref{fig:smeartopcharge} 
resembles the additive renormalization constant of 
$\langle Q_{\rm gluonic}^2 \rangle$ as function of the number of smearing 
steps~\footnote{The actual additive renormalization constant of the 
``fieldtheoretic topological density'' should of course be measured 
on the subsample of $Q_{\rm overlap}=0$ original Monte Carlo configurations.}.

\subsection{Monopoles removed}

Next, let us  describe what effect the removal of Abelian monopoles has on the 
P-vortex content in DMCG. The basic ensemble of configurations has been put 
into the MAG with the help of the simulated annealing method. 
The procedure of monopole removal is explained in Eqs. 
(\ref{eq:mremoved_Abelian}), (\ref{eq:premoved_Abelian}), 
(\ref{eq:mremoved_nonAbelian}) and (\ref{eq:premoved_nonAbelian}) in the Appendix. 

We have put these lattice fields into DMCG and viewed the corresponding 
P-vortex content. The P-vortex plaquette density 
$\rho_{\rm vort}~a^2 = 0.0231(4)$ of the original ensemble, 
{\it i.e.} the total area of vortex plaquettes relative to the total number of 
plaquettes in the lattice, is reduced to 
$\rho^{\prime}_{\rm vort}~a^2 = 0.0084(4)$ in the modified ensemble without 
monopoles. This amounts roughly to one third of the original density.

In the confinement phase, the contribution of the remaining center vortices to 
the quark-antiquark potential vanishes in the modified ensemble~\cite{mmp}. 
Here, the unphysical character of these left-over vortices is further 
elucidated by the smearing procedure as shown by the dashed line in 
Fig.~\ref{fig:smearmonvortdens} (b). The vortex density decays but not more 
than by an order of magnitude.

If configurations modified by monopole removal are gauge-fixed again to MAG
and Abelian projected, the number of monopole links that are then found is 
about one order of magnitude smaller than the number of monopoles defined 
for the original ensemble. The density of these artificial monopoles 
decreases extremely fast, by more than two orders of magnitude (!) per smearing 
step, as shown by the dashed curve in Fig.~\ref{fig:smearmonvortdens} (a).

The effect of monopole removal on the topological susceptibilty is much 
more pronounced. In Fig. \ref{fig:smeartopcharge} the dashed curve 
demonstrates that the gluonic estimator $Q_{\rm gluonic}^2$, which is 
immediately reduced before smearing, drops to zero within the first 
few smearing steps. The gluonic estimator 
is stronger suppressed by a factor of two to ten compared to the ensemble 
where P-vortices are removed. As we know, the fermionic topological charge 
is always $Q_{\rm overlap}=0$ as soon as monopoles have been removed.

For comparison we show in this 
Figure also $\langle Q_{\rm overlap}^2 \rangle$ obtained for the 
(unsmeared) equilibrium ensemble from the index of the overlap Dirac 
operator, which we have found~\cite{mmp1} to be stable under smearing.
The value of $\langle Q_{\rm overlap}^2 \rangle = 7.3 \pm 1.5$
is presented as the horizontal line.
The gluonic estimator $\langle Q_{\rm gluonic}^2 \rangle$ for the original
ensemble continues to grow with smearing (not shown in 
Fig.~\ref{fig:smeartopcharge} beyond $N_{\rm smear}=5$).
The same tendency to rise towards the final, non-vanishing gluonic estimator 
$\langle Q_{\rm gluonic}^2 \rangle$ we observe for 
the sample of photon-removed configurations. The fermionic value,
$\langle Q_{\rm overlap}^2 \rangle = 5.3 \pm 1.25$ is not yet reached. 

\section{Conclusions}
\label{sec:conclusions}

In this letter, in order to clarify the role of certain $SU(2)$ gauge field
excitations for the topological and chiral properties of the quenched 
ensemble of lattice fields, we have reconsidered the properties of the 
configurations at the deconfinement temperature $T=T_{\rm dec}$ after removing 
two typical non-perturbative degrees of freedom. The condensation of the 
corresponding gauge field fluctuations, Abelian monopoles and center vortices 
(more precisely, P-vortices), is popularly held responsible for quark confinement
at lower temperature. For our purpose we have applied special techniques of removal  
in order to study the effect on the complementary type of fluctuations and on 
the build-up of a topological charge and for the existence of a quark condensate.
We confirm that for overlap quarks and also at the deconfinement temperature 
the suppression of each of these non-perturbative degrees of freedom leads 
to a loss of the complementary type of fluctuations, of topological charge
and chiral quark condensate.
This effect is less pronounced for the loss of P-vortex density in the result 
of monopole suppression which might mean that the vortices, ceasing  
to percolate spatially at the deconfining temperature, become also decorrelated 
from the monopoles in the sense that they can exist also without monopoles. 
The converse is not true. The strength of this correlation is made visible by
applying smearing. 

Concerning the topological charge of a configuration we have the choice
between an (improved) gluonic definition of $Q_{\rm gluonic}$ and the index 
of the overlap Dirac operator. While the fermionic topological charge 
$Q_{\rm overlap}$ is immediately destroyed by removing monopoles or vortices, 
the gluonic topological susceptibility, in the absence of a true topological 
charge representing the additive renormalization for the topological 
susceptibility, rapidly drops to zero within a few smearing steps. 
For the unmodified ensemble, the actual topological charge  
$Q_{\rm overlap}$ (equal to plus or minus the number of zero modes)
is insensitive with respect to smearing~\cite{mmp1} 
even for more smearing steps than considered in this paper.
This smearing usually makes visible extended topological background 
excitations, as for example calorons or BPS monopoles (dyons).
These are suppressed as soon as Abelian monopoles or P-vortices are 
removed from the configurations such that no smearing can make them 
reappearing.

Finally we stress that, without any exception, quark condensation
is impossible without monopoles and P-vortices. 
This is seen by inspecting the spectra of the overlap Dirac operator. 
While differing in details, a gap is opended if monopoles or vortices 
are removed. No gap is opening and almost all zero modes are preserved 
if only the photon degrees of freedom are eliminated from the Abelian 
projection.

\section*{Appendix}
For a selfcontained readability of this paper 
we give the standard definitions for $SU(2)$ lattice gauge theory which
we have used in the studies described in the text. 
We perform our analyses in the Direct Maximal Center Gauges (DMCG). 
The DMCG in $SU(2)$ lattice gauge theory
is defined by the maximization of the functional

\begin{equation}
F^{\rm DMCG}_U(g) = \sum_{x,\mu} \left( \Tr~{}^gU_{x,\mu}\right)^2 \, ,
\label{eq:maxfunc_1}
\end{equation}
with respect to gauge transformations $g \in SU(2)$. 
$U_{x,\mu}= \{U_{x,\mu}^{jk}\}$ ($j,k=1,2$) is the lattice
gauge field and ${}^gU_{x,\mu}=g^{\dag}(x)U_{x,\mu}g(x+\hat{\mu})$ the gauge
transformed one.
Maximization of (\ref{eq:maxfunc_1}) fixes the gauge up to $Z(2)$ gauge
transformations, and the corresponding projected $Z(2)$ gauge field is
defined as:
\begin{equation}
Z_{x,\mu} = {\rm sign} \left( \Tr~{}^gU_{x,\mu} \right) \, .
\label{eq:Zdef}
\end{equation}
After this identification is made, one can make use of the remaining $Z(2)$ gauge
freedom in order to maximize the $Z(2)$ gauge functional
\begin{equation}
F^{\rm Z(2)}_Z(z) = \sum_{x,\mu} {}^zZ_{x,\mu} \,
\label{eq:maxfunc_2}
\end{equation}
with respect gauge transformations $z(x) \in Z(2)$, 
${}^zZ_{x,\mu}=z^{*}(x)Z_{x,\mu}z(x+\hat{\mu})$.
This is the $Z(2)$ equivalent of the Landau gauge. In distinction to
Ref.~\cite{deforcrand}, this final step is automatically understood here
before the vortex removal operation (to be defined below) is done.
The $Z(2)$ gauge variables are used to form $Z(2)$ plaquettes. 
The P-vortex surfaces are actually formed by plaquettes {\it dual} to the
negative plaquettes. 

The Maximally Abelian Gauge (MAG) is fixed 
by maximizing the functional
\begin{equation}
F^{\rm MAG}_U(g) = \sum_{x,\mu} \Tr~\left( {}^gU_{x,\mu}\sigma_3 ({}^gU_{x,\mu})^{\dag}
\sigma_3\right) \, ,
\label{eq:maxfunc_3}
\end{equation}
with respect to gauge transformations $g \in SU(2)$. The maximization fixes the
gauge up to $g \in U(1)$. Therefore, the following projection to an $U(1)$ gauge
field through the phase of the diagonal elements of the links,
$\theta_{x,\mu} = \arg ({}^gU_{x,\mu}^{11})$, 
is subject to a remaining $U(1)$ gauge freedom.
The non-Abelian link field is splitted according to
$U_{x,\mu}=u_{x,\mu} V_{x,\mu}$ in an Abelian (diagonal) part
$u_{x,\mu} = {\rm diag} \left\{ \exp( i \theta_{x,\mu} ),
                               \exp(- i \theta_{x,\mu} ) \right\}$ and a
coset part $V_{x,\mu} \in SU(2)/U(1)$, the latter representing non-diagonal gluons.

In order to fix the MAG and the DMCG we have created 10
randomly gauge transformed copies of the original gauge field configuration 
and applied the simulated annealing algorithm~\cite{simann} to find the optimal 
non-Abelian gauge transformation $g$. 

We have used the standard DeGrand--Toussaint definition~\cite{DGTmon} of
monopole currents defined by the phase $\theta_{x,\mu}$ of $u_{x,\mu}$.
The part of the Abelian gauge field originating from the monopoles is
\begin{equation}
\theta^{mon}_{x,\mu}  = -2 \pi \sum_{x^\prime} D(x-x^\prime)
\partial_{\nu}^{'} m_{x^{\prime},\nu\mu}\, .
\label{eq:monfield}
\end{equation}
Here $D(x)$ is the inverse lattice Laplacian, and $\partial_{\mu}^{'}$ is the
lattice backward derivative. 
The Dirac sheet variable, $m_{x,\mu\nu}$, is defined
as the integer multiple of $2\pi$ part of the plaquette angle $\theta_{x, \mu\nu}$,
whereas the reduced plaquette angle $\bar{\theta}_{x,\mu\nu}\in (-\pi,\pi]$ is the
fractional part:
$\theta_{x, \mu\nu} = 2\pi m_{x,\mu\nu} + \bar{\theta}_{x,\mu\nu}$.
The photon part is
\begin{equation}
\theta^{phot}_{x,\mu}  = \theta_{x,\mu} - \theta^{mon}_{x,\mu} \; .
\label{eq:photfield}
\end{equation}

The Abelian gauge field without monopole degrees of freedom is defined
as~\cite{miyamura_2}:
\begin{equation}
u_{x,\mu}^{{monopole\,\, removed}}=\left( u_{x,\mu}^{mon} \right)^{\dag}
u_{x,\mu}\, ,
\label{eq:mremoved_Abelian}
\end{equation}
where
$u_{x,\mu}^{mon} = {\rm diag} \left\{ \exp(  i \theta^{mon}_{x,\mu} ),
                                           \exp( -i \theta^{mon}_{x,\mu} ) \right\}$.
Correspondingly, the Abelian gauge field without the photon degrees of freedom 
is simply the monopole part 
\begin{equation}
u_{x,\mu}^{{photon\,\, removed}}= u_{x,\mu}^{mon} \; . 
\label{eq:premoved_Abelian}
\end{equation}

Upon multiplication with the coset field $V_{x,\mu}$, this holds also for
the non-Abelian links without monopoles
\begin{equation}
U_{x,\mu}^{{monopole\,\, removed}}=\left( u_{x,\mu}^{mon} \right)^{\dag} U_{x,\mu}\, 
\label{eq:mremoved_nonAbelian}
\end{equation}
and without photons
\begin{equation}
U_{x,\mu}^{{photon\,\, removed}}=\left( u_{x,\mu}^{phot} \right)^{\dag} U_{x,\mu}\, .
\label{eq:premoved_nonAbelian}
\end{equation}
Analogously the non-Abelian gauge fields without P-vortices are defined
as in Ref.~\cite{deforcrand}:
\begin{equation}
U_{x,\mu}^{{vortex\,\, removed}}=Z_{x,\mu} U_{x,\mu}\, ,
\label{eq:vremoved}
\end{equation}
where $Z_{x,\mu}$ is given by (\ref{eq:Zdef}).

Smearing is defined as an iterative field transformation with one step
\begin{equation}
U^{(n+1)}_{x,\mu} = {\cal P} \left\{ (1-\alpha) U^{(n)}_{x,\mu} 
+ \frac{\alpha}{6} \sum_{\nu,~\nu\ne\mu} \left( U^{(n)}_{x,\nu} 
                                                U^{(n)}_{x+\hat{\nu},\mu} 
                                                U^{(n)~\dagger}_{x+\hat{\mu},\nu} 
                                             +  U^{(n)~\dagger}_{x-\hat{\nu},\nu} 
				                U^{(n)}_{x-\hat{\nu},\mu}
					        U^{(n)}_{x-\hat{\nu}+\hat{\mu},\nu} \right)
						           \right\}
\end{equation}
where ${\cal P}$ denotes the projection to $SU(2)$.
This procedure has been used here to demonstrate the unphysical nature of 
monopoles, vortices and
gluonic topological charge apparently left over after the configurations 
have been modified by monopole/vortex removal. 
In the equilibrium ensemble, in contrast, these quantities are stable (up to 
changing renormalization) with respect to smearing, and the fermionic 
topological charge does not change at all. 
In this paper the smearing parameter has been set equal to $\alpha=0.5$.

The gluonic definition of the topological charge density is based on the 3-loop $O(a^4)$
improved field strength tensor~\cite{improved_field}
\begin{equation}
F_{\mu\nu}(x) = \left\{ \frac{1}{4} \sum_{\rm clover} \left( \frac{3}{2}  C^{(1)}_{\mu\nu}(x)
                                                           - \frac{3}{20} C^{(2)}_{\mu\nu}(x)
							   + \frac{1}{90} C^{(3)}_{\mu\nu}(x) \right)
							   \right\}_{\rm traceless} \; .
\end{equation}
where the ``clover'' average is taken over the four untraced, oriented Wilson loops $C^{(R)}(x)$
of size $R \times R$ in the $\mu\nu$ plane that are touching each other in site $x$ where they 
begin and end.  The gluonic topological charge is then
\begin{equation}
Q_{\rm gluonic} = \frac{1}{16\pi^2} \sum_x \sum_{\mu\nu\rho\lambda} 
\varepsilon_{\mu\nu\rho\sigma} \Tr~\left( F_{\mu\nu}(x)~F_{\rho\sigma}(x) \right) \; .
\end{equation}

\section*{Acknowledgements}

The work of ITEP group is partially supported by RFBR grants
RFBR 05-02-16306a and RFBR 07-02-00237a, by RFBR 06-02-04010 together 
with the DFG grant 436 RUS 113/739/2.
S.~M.~M. is partially supported by the INTAS YS fellowship 05-109-4821.
The work of E.-M.~I. is supported by DFG through the Forschergruppe 
FOR 465 (Mu932/2).

\end{document}